\documentclass[epj]{svjour}
\usepackage{graphicx}
\usepackage[latin1]{inputenc}

\begin{document}
\title{Exclusive measurement of quasi-free $\eta$-photoproduction
 from deuterium}
\author{J. Weiß\inst{1} \and 
P.~Achenbach\inst{2} \and
R.~Beck\inst{2} \and
V.~Hejny\inst{3} \and
V.~Kleber\inst{3} \and
M.~Kotulla\inst{1} \and
B.~Krusche\inst{4} \and
V.~Kuhr\inst{5} \and
R.~Leukel\inst{2} \and
V.~Metag\inst{1} \and
V.M.~Olmos de Léon\inst{1} \and
R.O.~Owens\inst{6} \and
F.~Rambo\inst{5} \and
A.~Schmidt\inst{2} \and
U.~Siodlaczek\inst{7} \and
H.~Ströher\inst{3} \and
F.~Wissmann\inst{5} \and
M.~Wolf\inst{1} 
}                     % Do not remove
\mail{Bernd.Krusche@unibas.ch} 
\institute{II.\ Physikalisches Institut,
Justus-Liebig-Universität Gießen, Heinrich-Buff-Ring 16, D-35392
Gießen \and
Institut
für Kernphysik,
Johannes-Gutenberg-Universität Mainz, Johannes-Joachim-Becher-Weg 45,
D-55099 Mainz 
\and
Institut für Kernphysik, Forschungszentrum Jülich GmbH,
D-52425 Jülich 
\and
Department für Physik und Astronomie, Universität  Basel,
Klingelbergstrasse 82, CH-4056 Basel
\and
II.\ Physikalisches Institut,
Georg-August-Universität Göttingen, Bunsenstr.\ 7-9, D-37073
Göttingen 
\and
Department of Physics and Astronomy, University of Glasgow, Glasgow
G128QQ, UK
\and
Physikalisches Institut,
Eberhard-Karls-Universität Tübingen, Auf der Morgenstelle 14, D-72076
Tübingen
}
\date{Received: date / Revised version: date}
\abstract{
Quasi-free photoproduction of $\eta$-mesons from the deuteron has been measured 
at the tagged photon facility of the Mainz microtron MAMI with the photon 
spectrometer TAPS for incident photon energies from the production threshold 
at 630 MeV up to 820 MeV. In a fully exclusive measurement $\eta$-mesons and 
recoil nucleons were detected in coincidence. At incident photon energies above
the production threshold on the free nucleon, where final state interaction 
effects are negligible, an almost constant ratio of 
$\sigma_n/\sigma_p=0.66\pm0.10$ was found. At lower incident photon energies 
the ratio rises due to re-scattering effects. The average ratio agrees with the 
value extracted from a comparison of the inclusive $d(\gamma ,\eta)X$ cross 
section to the free proton cross section via a participant - spectator model 
for the deuteron cross section ($\sigma_n/\sigma_p=0.67\pm0.07$). The angular 
dependence of the ratio agrees with the expected small deviation of neutron 
and proton cross section from isotropy due to the influence of the 
D$_{13}$(1520) resonance. The energy dependence of the ratio in the excitation 
energy range of the S$_{11}$(1535) resonance disfavors model predictions which 
try to explain this resonance as $K\Sigma$-bound state.
\PACS{
	{13.60.Le}{Meson production} \and
	{25.20.Lj}{photoproduction reactions}
     } % end of PACS codes
} %end of abstract
\maketitle
\section{Introduction}
\label{sec:intro}
The first excited state of the nucleon with the quantum numbers 
$I(J^P)=1/2(1/2^-)$, the so-called S$_{11}(1535)$ resonance has very interesting
properties and its structure was recently intensely debated in the literature
\cite{Glozman,Kaiser_1,Kaiser_2,Bijker}. The discussion centered around the 
unusual large decay branching ratio ($b_{\eta}\approx$ 50\% \cite{PDG}) 
of this resonance into the $N\eta$ channel. The much weaker coupling of
the close-by P$_{11}(1440)$ and D$_{13}(1520)$ resonances to the 
$N\eta$ channel can be explained by simple phase space arguments since they 
need to decay with relative orbital angular momentum $l=1,2$ which is  
suppressed close to threshold. However, also the second S$_{11}$-resonance at 
1655 MeV has only a branching ratio into $N\eta$ of 1 \% or less. 
Glozmann and Riska \cite{Glozman} have argued, that these decay patterns
can be understood as a isospin selection rule in their quark model with 
quark-diquark clusterization of the nucleon. Their diquark clusters in the 
nucleon ground state and in the S$_{11}(1535)$ have both $I=0$ but the 
diquark  cluster in the S$_{11}(1650)$ has $I=1$. Consequently, the decay via 
the isoscalar $\eta$-meson is strongly suppressed for the second S$_{11}$.
Other work even questioned the very nature of the S$_{11}(1535)$ as a nucleon
resonance.  Kaiser et al. \cite{Kaiser_1,Kaiser_2} treated the resonance as a
quasi-bound $K\Sigma$-state and Bijker et al. \cite{Bijker}, failing to
explain the S$_{11}(1535)$ properties in their model, suggested a quasi-bound
$N\eta$ (penta-quark) state. If such suggestions are correct, the immediate
question would be where are the regular low lying three quark states predicted 
in the quark model with the quantum numbers 1/2$^-$,1/2?
 
The precise experimental determination of the properties of this state 
is essential in order to resolve the problem. From the experimental point of
view the strong coupling to the $N\eta$-channel is advantageous
since it allows an almost background free investigation of this resonance. 
Precise experiments measuring angular distributions
\cite{Krusche_1}, the photon beam asymmetry \cite{Ajaka} and the target 
asymmetry \cite{Bock} for $\eta$-photoproduction and the dependence of the 
electromagnetic helicity coupling on the momentum transfer in
$\eta$-electroproduction \cite{Thompson} have been recently reported for the
proton. However, photoproduction from the proton alone gives no information 
about the isospin structure of the electromagnetic excitation of the resonance.
It is thus desirable to study the reaction also on the neutron, but due to 
the non-availability of free neutron targets such experiments must rely on 
meson photoproduction from light nuclei. Due to its small binding energy and 
the comparatively well understood nuclear structure the deuteron is the 
obvious choice as target nucleus.

Previously, quasifree $\eta$-photoproduction from the deuteron was measured in
two experiments \cite{Krusche_2,Hoffmann} carried out at the Mainz MAMI- and 
Bonn ELSA-accelerators. In the first experiment \cite{Krusche_2} only the 
$\eta$-mesons were detected via their two-photon decay channel. The
neutron - proton cross section ratio was determined from the comparison of the
inclusive cross section to the free proton cross section in the framework of a
participant - spectator model. In the second experiment \cite{Hoffmann}
recoil protons and neutrons were measured, but the $\eta$-mesons were only 
identified via a missing mass analysis of the recoil nucleons. 
In the present experiment $\eta$-mesons were unambiguously identified in an
invariant mass analysis of their decay photons and the recoil nucleons were
detected in coincidence. This allows a very clean comparison of the proton
and neutron cross sections measured under identical conditions from the 
nucleons bound in the deuteron. The kinematic over-determination furthermore 
allows the test of the participant - spectator assumption. At the same time
the cross section of the inclusive reaction $d(\gamma, \eta)X$ was re-measured
with much higher statistical accuracy.

\section{Experimental setup}
\label{sec:experiment}

The experiment was carried out using the Glasgow tagged photon facility
\cite{Anthony} installed at the Mainz microtron MAMI \cite{Walcher,Ahrens} 
with the TAPS detector \cite{Novotny,Gabler}. 

The quasi-monochromatic photon beam covered the energy range from
580 - 820 MeV with a resolution of 1 - 2 MeV and an average photon flux of
$\approx$ 250 kHz/MeV. The liquid deuterium target of 10~cm length was 
mounted in a vacuum chamber of 90~cm diameter, constructed from carbon fiber.
The thin wall thickness (4~mm) of the chamber ensured that scattering and 
energy loss of the recoil nucleons were minimized.
The photons from the $2\gamma$-decay of the $\eta$-mesons and the recoil
nucleons were detected with the electromagnetic calorimeter TAPS.
The device consists of 504 BaF$_2$ crystals and is optimized 
for the detection of photons via electromagnetic showers but has also 
excellent particle detection capabilities. Each $\mathrm{BaF}_2$ scintillator
module is of hexagonal shape, 25~cm long (corresponding to 12 radiation 
lengths), and has an inner diameter of 5.9 cm. The modules were arranged in six
blocks with 8$\times$8 $\mathrm{BaF}_2$ crystals each, placed at polar angles 
of $\pm 50^\circ$,$\pm 100^\circ$, and $\pm 150^\circ$ in a plane around the 
scattering chamber at a distance of 54 cm from the target center. A hexagonally
shaped forward wall of 120 crystals was positioned at $0^\circ$ around the 
beam pipe at a distance of 1 m from the target. The BaF$_2$-modules of the 
blocks were equipped with individually read-out plastic scintillators (NE102) 
at their front sides for the  identification of charged particles. The modules 
of the forward wall were operated in phoswich mode, i.e. the light from each 
BaF$_2$-module and the corresponding plastic scintillator was read out by the 
same photo-multiplier tube. These modules effectively worked as 
$\Delta$E-E-telescopes and thus allowed the separate identification of photons,
recoil protons and neutrons. A detailed description of the experimental setup 
is given in \cite{Hejny}, where the results for quasifree $\eta$-photoproduction
from $^4He$ are reported.

\section{Data analysis}
\label{sec:analysis}

The separation of photons from recoil particles was achieved with a 
time-of-flight measurement with typically 500 ps resolution (FWHM), the 
pulse-shape discrimination capability of the BaF$_2$-scintillators, and the 
charged particle veto detectors. The energies and impact points of the photons 
were determined in the usual way (see e.g. \cite{Hejny}) from the energies 
deposited by the electromagnetic showers in clusters of BaF$_2$ scintillator 
modules. The energy calibration of the individual modules was done with the 
help of cosmic radiation.
The $\eta$-mesons were then identified via their two photon decay, which has a
branching ratio of 39.2\%, with a standard invariant mass analysis of
coincident photon pairs using:
\begin{equation}
  \label{eq:etot}
  m_{\mathrm{inv}}^2  = ( E_1 + E_2 )^2 - (\vec{p}_1 + \vec{p}_2)^2 
          =  2 E_1 E_2 (1-\cos \Phi_{12}) 
\end{equation}
where $E_i$ denotes the energies, $\vec{p}_i$ the momenta of the two photons
and $\Phi_{12}$ the relative angle between them (units with $\hbar =c=1$ are
used throughout the paper). An invariant mass
resolution for the $\eta$ of $\approx$ 60 MeV (FWHM) was achieved (see
\cite{Hejny}). The data sample with an identified $\eta$-meson was used to
extract the angular distributions and the total cross section of inclusive 
$\eta$-photoproduction from the deuteron. The energy and angle dependent
detection efficiency of the TAPS detector for the 2$\gamma$-decay of 
$\eta$-mesons was determined as described in \cite{Weiss_2,Krusche_3} with a 
Monte Carlo simulation using the GEANT code \cite{Geant}. 
It is important to note that no model assumptions about the non-trivial
energy and angular distributions of the $\eta$-mesons in the three body final
state were needed for the modeling of the detection efficiency. This is 
possible because in the investigated energy range the detector has a 
non-vanishing detection probability for all kinematically possible combinations
of the meson lab angle and energy.
The absolute normalization of the cross section was obtained in the usual way 
from the target thickness and the photon flux \cite{Weiss_2,Krusche_3}. 

The recoil nucleons were detected in the forward wall which subtended
laboratory angles between 5$^{\circ}$ and 20$^{\circ}$. Due to the Lorentz 
boost, only a few nucleons were seen in the side-ward blocks which covered 
lab angles larger than 35$^{\circ}$. 
The recoil protons and neutrons were identified by their pulse shape signature 
and their time-of-flight (see \cite{Hejny}). 
The standard energy calibration of TAPS, which is done with cosmic muons, is 
only valid for photons since the energy response of the BaF$_2$-plastic 
phoswich telescopes is different for particles. Furthermore charged particles 
like protons undergo some energy loss in the target and the wall of the 
scattering chamber. The energy information of the protons 
was re-calibrated using the kinematic over-determination of the reaction 
when the $\eta$ is detected in coincidence with the proton.
This is possible because in the energy range of interest $np\eta$ is the only 
final state with proton and $\eta$-meson. 
In this case the only missing particle is the neutron and the energy of the 
proton can be varied until the missing mass of the reaction (see below)
equals the neutron mass. This correction was not done event-by-event but an
average correction as function of the proton energy obtained from the
cosmic calibration was determined.
Since there is no direct correlation of the energy response of the detector 
and the kinetic energies of the neutrons, the neutron energies were determined 
from the time-of-flight measurement. The energy resolution achieved by this
procedures can be judged from the missing mass spectra where the mass of the non
observed nucleon is derived from the energy of the incident photon and
the momenta of the $\eta$-meson and the detected nucleon using:
\begin{eqnarray}
  \Delta E & = & (E_d+E_{\gamma})-(E_{\eta}+E_p) \label{eq:ebilanz} \\
  \Delta \vec{p} & = & \vec{p}_{\gamma}-(\vec{p}_{\eta}+\vec{p}_p)
   \label{eq:pbilanz}\\
  m_\mathrm{miss}^2 & = & \Delta E^2 - \Delta \vec{p}^2   \label{eq:mmiss}
\end{eqnarray}
where $E_d$, $E_{\gamma}$, $E_{\eta}$ and $E_p$ are the total energies of
the incident deuteron and photon and the detected $\eta$-meson and participant
nucleon and $\vec{p}_{\gamma}$, $\vec{p}_{\eta}$, $\vec{p}_p$ are the respective
momenta .
\begin{figure}[b]
\begin{center}
\includegraphics[width=\columnwidth]{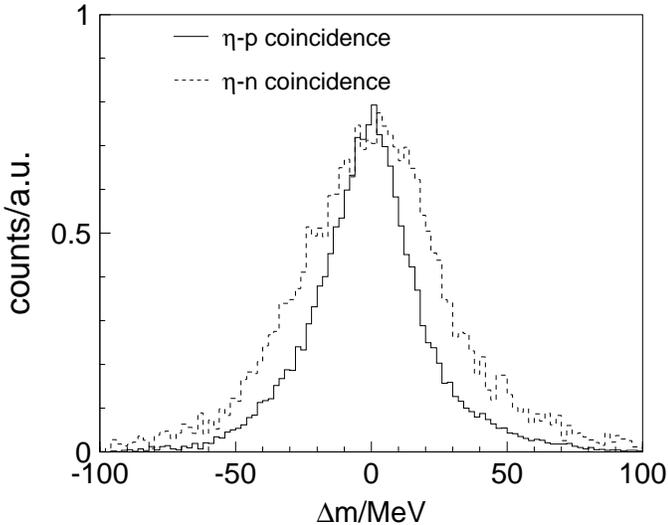}
\end{center}
\caption{Missing mass of the coincident $\eta$ nucleon system. The
expected mass value is subtracted.}
\label{fig:1}
\end{figure} 

The expected value for the missing mass (i.e. proton or neutron mass) is
subtracted from $m_\mathrm{miss}$, so that the resulting distributions 
for $\Delta m=m_{miss}-m_N$ peak at zero, which in case of the $\eta$-proton 
coincidence is enforced by the calibration procedure. 
The result of this analysis is shown in fig. \ref{fig:1}. 
The resolution for the $\eta$-neutron coincidence is not as good as for the
$\eta$-proton coincidence due to the limited time-of-flight resolution
(500 ps FWHM, 1 m flight path).

As discussed above, the detection efficiency of the $\eta$-mesons was simulated
with the GEANT package, but the simulation of neutron and proton  
efficiencies is much less precise. The recoil particle detection efficiencies
were thus experimentally determined with the help of reactions where the
nucleons are 'tagged' by the detection of other reaction products. The reaction
$p(\gamma,\pi^0)p$ with the detection of the $\pi^o$ was used for the 
proton efficiency and the reaction $p(\gamma ,\pi^0\pi^+)n$ with detection of
the $\pi^0$- and $\pi^+$-mesons for the neutron efficiency. In both cases the 
momentum of the recoil nucleon was determined from the incident photon energy 
and the measured momenta of the pions and the efficiencies were constructed 
from the ratios of the event numbers with and without additional detection of 
the recoil nucleon. The procedure is described in detail in \cite{Hejny}.

The acceptance of the TAPS detector for the two-photon decay of $\eta$-mesons
covers the full polar angular range. However, recoil nucleons were only
detected for lab angles between 5$^{\circ}$ and 20$^{\circ}$ and recoil
protons with kinetic energies below $\approx$ 60 MeV did not reach the
detector. Since an extrapolation to the full angular and energy range of the 
recoil nucleons would be model dependent, the exclusive cross sections
are only given in the TAPS acceptance. This means, that only events with
recoil nucleons in the restricted angular and energy ranges were considered 
(neutrons with energies below the proton detection threshold were removed)
and corrected for the TAPS detection efficiencies for $\eta$-mesons, neutrons
and protons discussed above. 
In this way, no total exclusive cross sections were obtained, however the
cross sections within the TAPS acceptance can be used to extract the neutron -
proton cross section ratio. Here one has to keep in mind that due to the 
Lorentz boost a large fraction of the recoil nucleons is emitted into the 
angular range covered by the forward wall (see \cite{Hejny}). Furthermore is 
important, that as shown below, the cross section ratio is not strongly angular
dependent and that due to the influence of Fermi motion a modest restriction 
of the angular range of the recoil nucleons does not restrict the angular range
of the coincident $\eta$-mesons. 

\section{Results and discussion}
\label{sec:result}

\subsection{Inclusive $\eta$ cross section}
\label{sec:etaincl}

\begin{figure}[hbt]
  \includegraphics[width=\columnwidth]{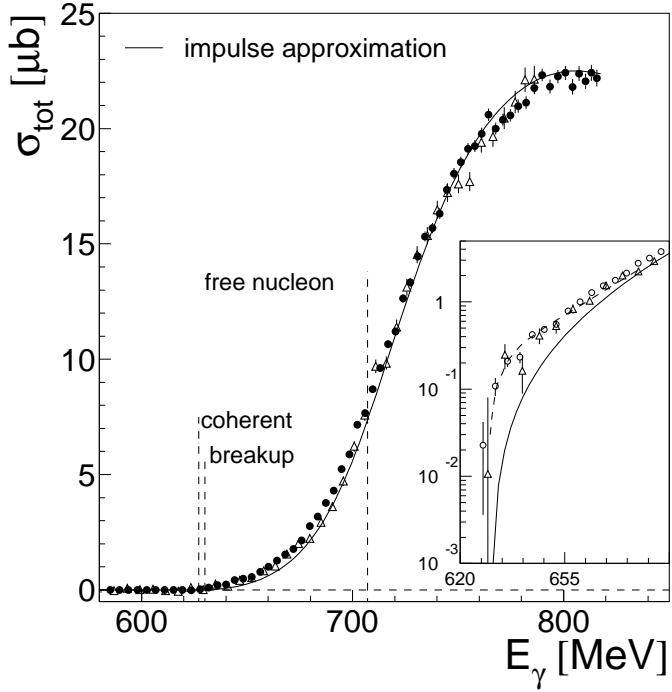}
\caption{Inclusive $\eta$ photoproduction cross section from the
deuteron. Circles: present experiment, triangles: ref. \cite{Krusche_2}. 
The dashed lines give the deuteron coherent, the deuteron breakup
and the free nucleon production thresholds. The solid curve is a fit of the 
impulse approximation model to the data (see text). Insert: threshold region.
The dashed curve is the result of a model including FSI effects 
\cite{Sibirtsev_2}}.
\label{fig:etaincl}       % Give a unique label
\end{figure}

Eta-photoproduction from the neutron can be measured via the 
$d(\gamma ,\eta)np$ reaction in so-called quasi-free kinematic where the 
$\eta$-meson is produced on one nucleon, called the participant, while the 
other nucleon can be regarded as a spectator.
The total inclusive cross section for $\eta$-photoproduction from the deuteron
is shown in fig. \ref{fig:etaincl}. The data agrees with our previous result
\cite{Krusche_2}, but has much higher statistical quality and extends to 
somewhat higher incident photon energies so that the resonance position of the
S$_{11}$-state ($W_R\approx$ 1540 MeV, $E_{\gamma}\approx$ 795 MeV) is covered.
The extraction of the cross section ratio $\sigma_n /\sigma_p$ from this data
relies on the participant - spectator approximation as in
\cite{Krusche_2,Hejny}. A Breit-Wigner curve fitted to the cross section of the
$p(\gamma ,\eta)p$ reaction was folded with the momentum distribution of the
bound nucleons calculated from the deuteron wave function \cite{Lacombe}.
The result was adjusted to the measured inclusive $d(\gamma ,\eta)X$ cross
section with a constant factor representing an energy independent 
$\sigma_n /\sigma_p$ ratio. The result of this fit is also shown in fig.
\ref{fig:etaincl}. At energies above the $\eta$-production threshold on the free
nucleon the data is very well reproduced with the constant ratio 
$\sigma_n/\sigma_p = 0.67\pm0.005\pm0.07$, where the first error is statistical
and the second comes from the systematic uncertainties of the proton and
deuteron cross sections.  At lower incident photon energies the data is 
underestimated by this fit. This is however expected since it is known
\cite{Fix,Sibirtsev,Weiss_2,Sibirtsev_2} that in this energy range the 
cross section from the deuteron is enhanced by final state 
interaction effects. The dashed curve in the insert of fig. \ref{fig:etaincl} 
shows the result of a calculation with the same ratio of $\sigma_n/\sigma_p$ 
which includes $NN$- and $\eta N$-FSI effects \cite{Sibirtsev_2} and is in 
good agreement with the data. 

\newpage

\subsection{The participant spectator model}

The extraction of the cross section ratio $\sigma_n /\sigma_p$ from inclusive 
data where only the $\eta$-meson is detected, relies on two assumptions. The 
contribution from the coherent reaction $d(\gamma ,\eta)d$ must be negligible,
which is true for the deuteron in the energy range of interest
\cite{Krusche_2,Hoffmann,Weiss}. The other assumption is the validity of the 
spectator - participant model which is used to extract the cross section ratio
from a comparison of the inclusive deuteron data to the cross section on the 
free proton folded with the momentum distribution of the nucleons bound in the 
deuteron \cite{Krusche_2,Hejny}.   
This means in particular that the cross section from the bound nucleons must
not be affected by off-shell effects beyond nuclear Fermi motion and
re-scattering contributions must be negligible. Such effects can be much better
controlled when the recoil nucleon is detected in coincidence with the
$\eta$-meson. In this case, one can directly compare the cross section of the
bound proton to the bound neutron and furthermore one can select kinematic
conditions close to the ideal quasi-free situation.

\begin{figure}[hbt]
  \includegraphics[width=\columnwidth]{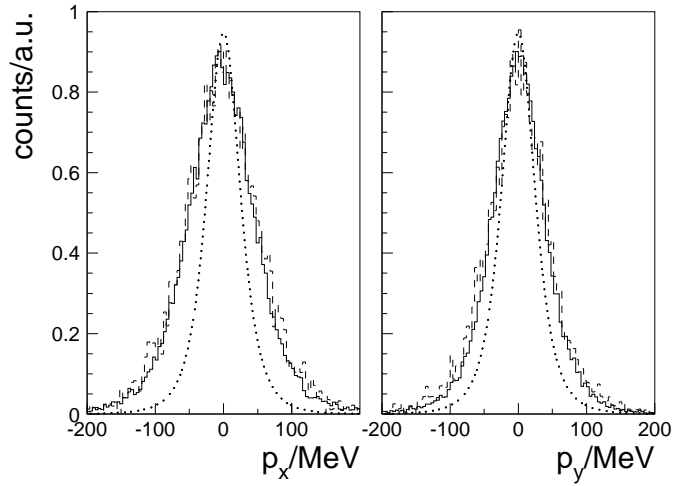}
\caption{x (left) and y (right) momentum distributions of
reconstructed proton (solid) and neutron (dashed), respectively.
Dotted curves: momentum distribution of the bound nucleons
calculated from the deuteron wave function without taking resolution effects 
into account.}
\label{fig:pxy}       % Give a unique label
\end{figure}

The lab momentum of the non-detected spectator nucleon was constructed from 
the measured momenta of the $\eta$-meson and the participant nucleon according 
to eq. \ref{eq:pbilanz}. The x- and y-components (perpendicular to the photon
beam) of the momentum are shown in fig. \ref{fig:pxy}.
\begin{figure}[t]
  \includegraphics[width=\columnwidth]{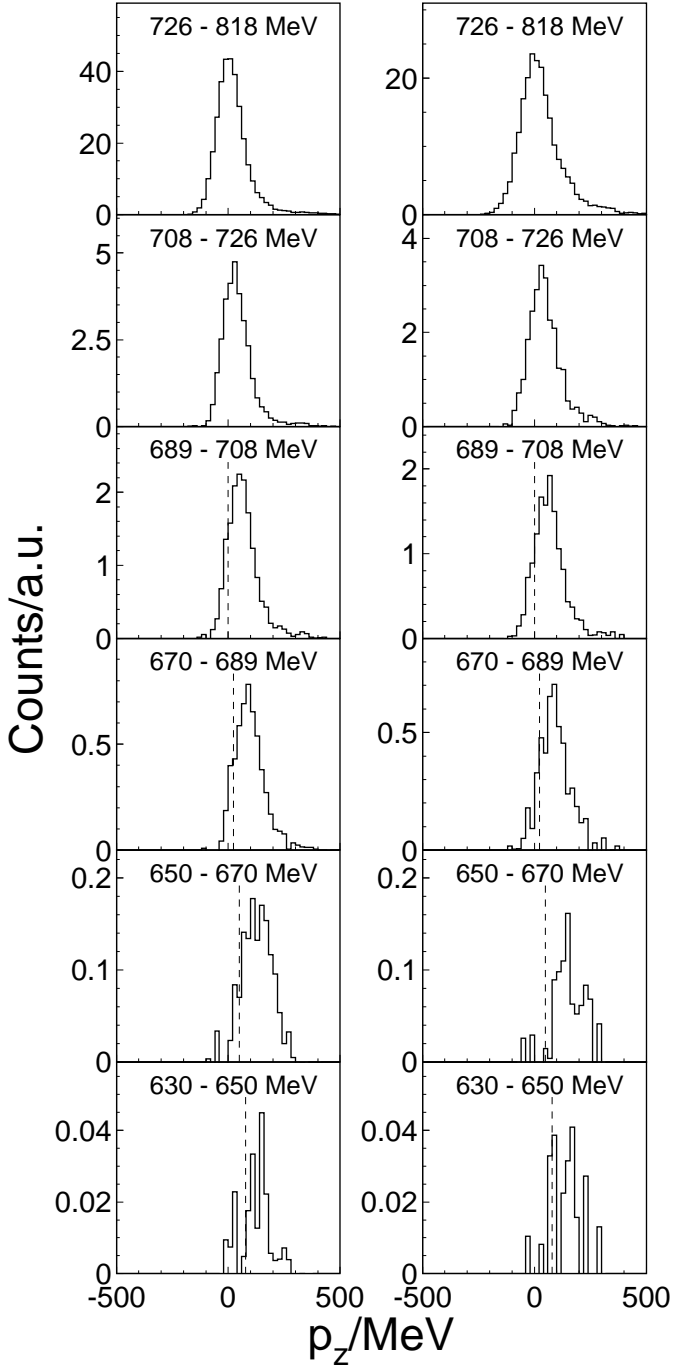}
\caption{Reconstructed z-component of the momentum of the spectator nucleon 
(left: reconstructed proton, i.e. participating neutron, right: reconstructed
neutron, i.e. participating proton). Dashed lines give the minimum
required momentum to produce the $\eta$ meson for that particular
photon beam energy.}
\label{fig:pz}       % Give a unique label
\end{figure}
As expected the distributions are centered around zero. 
In the participant - spectator 
approximation the spectator nucleon picks up the momentum $-\vec{p}_F$ where 
$\vec{p}_F$ is the Fermi momentum of the participant nucleon which is 
symmetrically distributed around zero. The momentum distribution of the bound 
nucleons calculated from the deuteron wave function \cite{Lacombe} is shown for
comparison. The difference between the measured distributions and the prediction
from the wave function is mainly due to instrumental resolution effects.

The z-component (parallel to the photon beam) of the spectator 
nucleons is shown in fig. \ref{fig:pz} for different bins of incident photon
energy. In this case the distribution is approximately symmetric around zero
for high incident photon energies, but the situation changes for low incident
photon energies. At energies below the production threshold on the free nucleon
($\approx$ 707 MeV) the reaction is only possible for participant nucleons
with a z-component of the Fermi momentum anti-parallel to the photon beam.
Consequently, spectator momenta in photon beam direction are selected.
Very close to the threshold for coherent $\eta$-production from the deuteron
($\approx$ 628 MeV) energy and momentum conservation enforce that 
spectator and participant nucleon have only a small relative momentum 
in the final state. This means that for very low incident photon energies
no experimental separation of 'participants' and 'spectators' is possible and 
the probability for final state interaction between the nucleons becomes 
large so that the participant - spectator approximation breaks down.

In the ideal case where re-scattering effects etc. can be neglected, the 
measured quasi-free cross sections correspond to the elementary cross
sections on the nucleon folded with the momentum distributions of the bound
nucleons. However, due to the kinematic over-determination of the exclusive
reaction it is also possible to reconstruct event-by-event the total cm energy
of the participant nucleon - $\eta$ system via:
\begin{equation}
\label{eq:s_eff}
s=(E_{\eta}+E_p)^2-(\vec{p}_{\eta}+\vec{p}_{p})^2
\end{equation}
and to deduce the effective incident photon energy $E_{\gamma}^{\star}$ which
would lead to the same $\sqrt{s}$ for an incident nucleon at rest:
\begin{equation}
\label{eq:e_eff}
E_{\gamma}^{\star}=
\frac{s-m^2_p}{2m_p}\;.
\end{equation}   
In this way it is not only possible to eliminate the Fermi smearing but one can
even extract a cross section ratio for values of $\sqrt{s}$ which are larger
than $\sqrt{s}$ for a given incident photon energy on a nucleon at rest.
  
\subsection{Neutron-proton cross section ratio}
\label{sec:ratio}

\begin{figure}[th]
  \includegraphics[width=0.95\columnwidth]{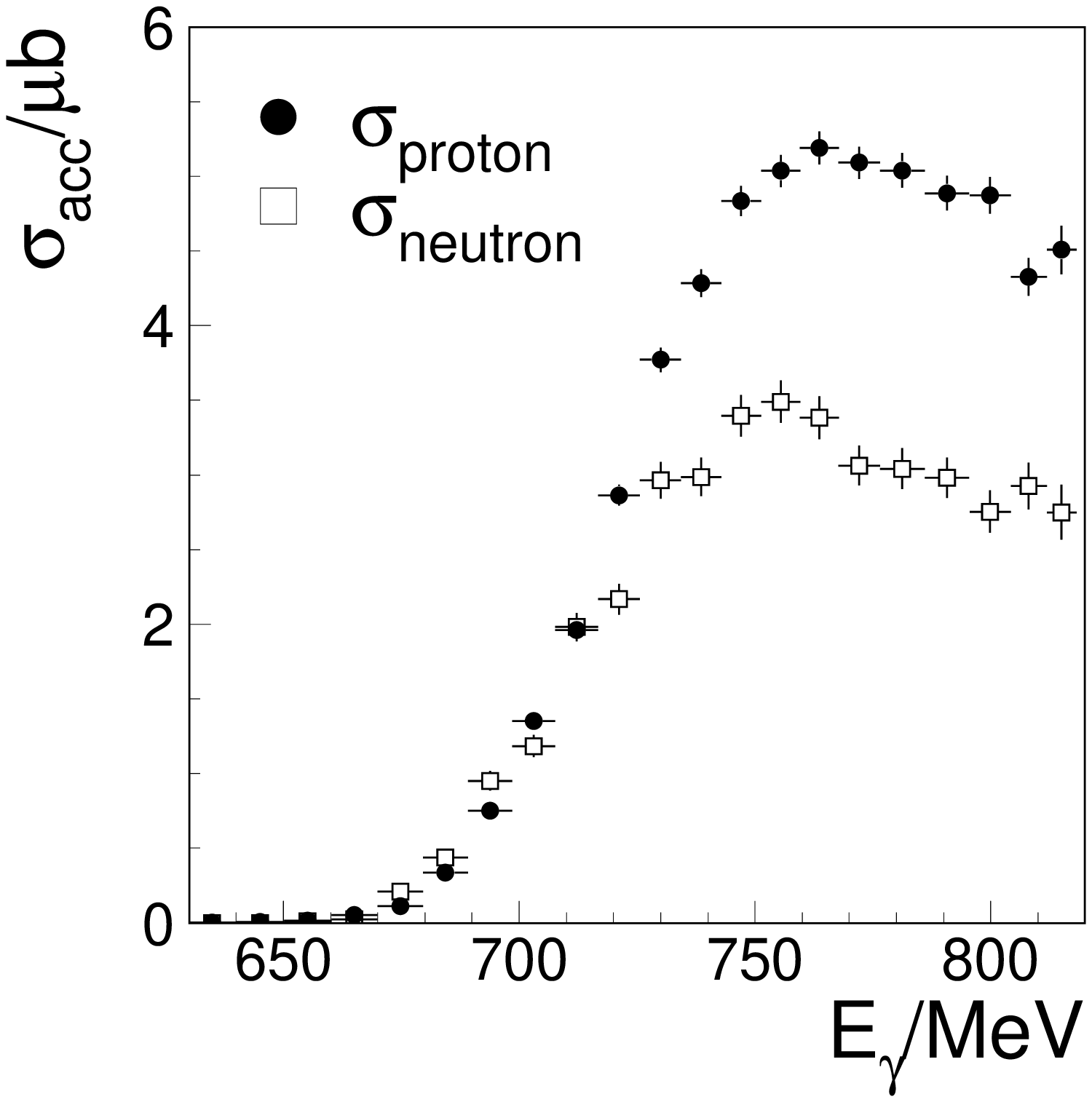}
    \includegraphics[width=0.95\columnwidth]{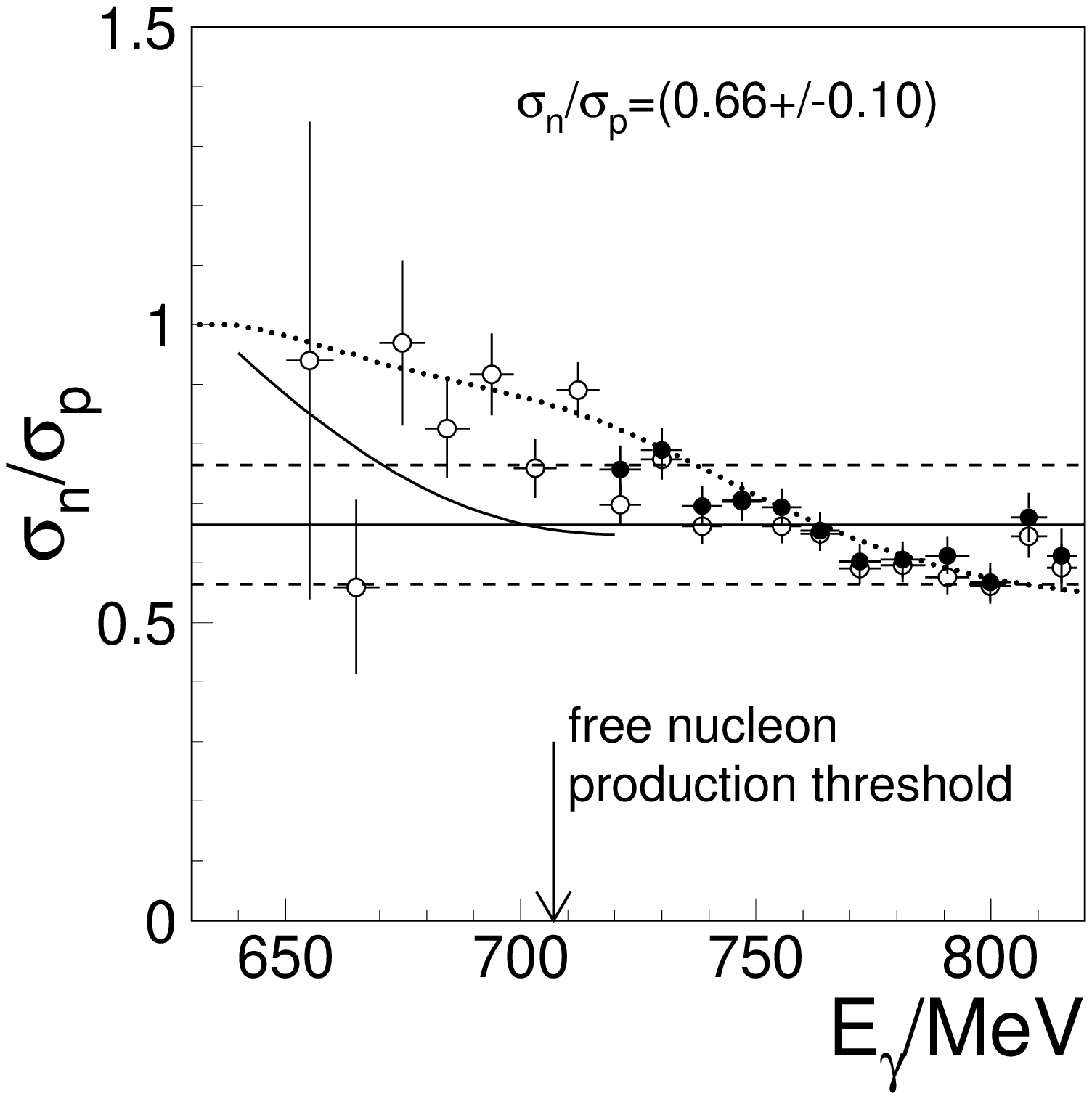}
\caption{Upper part: cross section for coincident measurement of $\eta$
meson and nucleon (the geometrical acceptance for the recoil nucleons
is not corrected).
Lower part: ratio of $\sigma_n/\sigma_p$ without ({\Large $\circ$}) and with 
({\Large $\bullet$}) cut on z-momentum of reconstructed nucleon to select
participant nucleons (see text). The solid line is a calculation of
Fix et al. \cite{Fix}. The dotted line is the ratio of Fermi smeared
proton and neutron cross sections corresponding to the neutron - proton
ratio as function of equivalent photon energies (see text below).} 
\label{fig:proneu}       % Give a unique label
\end{figure}
    
The exclusive cross sections in the TAPS acceptance and the neutron - proton
ratio as function of the beam energy are summarized in fig. \ref{fig:proneu}. 
The ratio over the full range of incident photon energies was obtained with 
no further cuts on the reaction kinematics. In addition, above the production 
threshold on the free nucleon a ratio with a cut on the z-component of the 
spectator nucleon around zero ($\pm 1.5 \sigma$ of a Gaussian fitted to the 
$p_z$ distribution in the highest photon energy bin of 726--818~MeV) is shown 
in the figure. This cut removes events in which due to large Fermi momenta 
spectator nucleons have been detected in TAPS and falsely interpreted as 
participant nucleons. However the effect of the cut is small since at the 
higher incident photon energies participant and spectator nucleons are 
well distinguished.

The average of the ratio above the production threshold on the free nucleon
is 0.66$\pm$0.01 (statistical error). However, even in this 
energy range the data seem to show some dependence on the incident photon 
energy. Taking into account this possible variation a conservative
estimate of $\sigma_n/\sigma_p = (0.66\pm0.10)$ is derived for photon energies
above the free nucleon production threshold.

The cross section ratio rises significantly at lower incident photon energies
and approaches unity. Such a behavior was predicted by Fix and Arenh\"ovel
\cite{Fix} with a model calculation which takes into account $NN$ and $\eta N$
re-scattering. This rise therefore does not reflect an energy dependence of 
the elementary cross section ratio but is a pure nuclear effect. 
Their prediction for a fixed nucleon lab polar angle of 4$^{\circ}$ is 
compared to the data in fig. \ref{fig:proneu}. Here one should keep in mind 
that in the immediate vicinity of the threshold the kinematic separation 
between spectator and participant nucleons breaks down. This means that very 
close to threshold the experimentally observed cross section ratio will 
approach unity in any case.

Inclusive and exclusive total cross section data allow only the extraction of
the ratio of the absolute values of $A_{1/2}^n$, $A_{1/2}^p$ but not of the
relative sign which is related to the isospin structure of the helicity
amplitudes via:
\begin{eqnarray}
  \label{eq:isospin_1}
  |A^p_{1/2}|^2 & = &|A^{IS}_{1/2} + A^{IV}_{1/2}|^2\\
  |A^n_{1/2}|^2 & = & |A^{IS}_{1/2} - A^{IV}_{1/2}|^2\nonumber\; 
\end{eqnarray}
where $A^{IS}_{1/2}$ denotes the isoscalar and $A^{IV}_{1/2}$ the isovector 
part of the helicity amplitude for the proton ($A^p_{1/2}$) and the neutron
($A^n_{1/2}$), respectively. Since the ratio of the absolute values is close to 
0.8, either the isovector or the isoscalar part must be strongly dominant
and the relative sign between $A_{1/2}^n$ and $A_{1/2}^p$ determines which one.
This question has been investigated with the help of coherent
$\eta$-photoproduction from the deuteron \cite{Krusche_2,Hoffmann,Weiss}.
In this case only the isoscalar amplitude contributes and the observed 
very small cross sections have been taken as evidence for a dominant isoscalar
contribution. 

A different possibility to determine the relative sign originates
from interference terms in the angular distributions of quasifree 
$\eta$-production from the deuteron. This can be seen in the following way.
Due to the strong dominance of the S$_{11}$-resonance in $\eta$-photoproduction,
the cross section can be expressed in good approximation by s-wave multipoles 
and interference terms of the s-wave with other multipoles 
\cite{Krusche_1,Tiator_99} which gives:
\begin{eqnarray}
\frac{d\sigma}{d\Omega}=&&\frac{q_{\eta}^{\star}}{k^{\star}}
[E_{o^+}^{2} - \mbox{Re}(E_{o^+}^{\star}(E_{2^-}-3M_{2^-}))\nonumber\\
&&+2\mbox{Re}(E_{o^+}^{\star}(3E_{1^+}+M_{1^+}-M_{1^-}))cos(\Theta^{\star})\nonumber\\
&&+3\mbox{Re}(E_{o^+}^{\star}(E_{2^-}-3M_{2^-}))cos^{2}(\Theta^{\star})]\;\;\;.
\end{eqnarray}
It has been shown \cite{Krusche_1,Tiator_99}, that in the excitation energy 
range of the S$_{11}$(1535), the cross section is dominated by the constant
term, the $cos$-term is negligible, and the angular dependence results
entirely from the $cos^2$-term. The responsible amplitude ${\cal{I}}$:
\begin{equation}
{\cal{I}} = Re(E_{o^+}^{\star}(E_{2^-}-3M_{2^-}))
\end{equation}
results from an interference of the S$_{11}$ excitation ($E_{o^+}$-multipole) 
with the D$_{13}$ excitation ($E_{2-}$-, $M_{2-}$-multipoles)
and can be rewritten in terms of helicity elements as:
\begin{equation}
{\cal{I}} \propto 
\mbox{Re}(A_{o+}^{\star}A_{2-})\;,
\end{equation}
in the usual notation.
Close to the resonance positions of the S$_{11}$ and the D$_{13}$, the product
of the real parts of the amplitudes is negligible compared to the product of 
the imaginary parts so that:
\begin{equation}
{\cal{I}} \propto 
\mbox{Im}(A_{o+})\mbox{Im}(A_{2-})\;\;.
\end{equation}
 In the vicinity of the two resonances this can be approximated by:
\begin{equation}
{\cal{I}} \propto
A_{1/2}^{N}(S_{11}) A_{1/2}^{N}(D_{13})
\times {\cal{F}}(E_{\gamma})
\end{equation}
where ${\cal{F}}(E_{\gamma})$ is the same function for proton and neutron and
$A_{1/2}^{N}$ are the helicity 1/2 couplings for the resonance excitations
on proton ($N=p$) and neutron ($N=n$). 
Therefore the angular distributions can be approximated by:
\begin{eqnarray}
\label{eq:dsapp}
\left(\frac{d\sigma}{d\Omega}\right)_{N} & \propto & (A_{1/2}^N(S_{11}))^2 +\\
& & {\cal{G}}(E_{\gamma})A_{1/2}^N(S_{11})A_{1/2}^N(D_{13})(3cos^2(\Theta)-1)
\;\;.\nonumber
\end{eqnarray}
\begin{figure}[t]
  \includegraphics[width=\columnwidth]{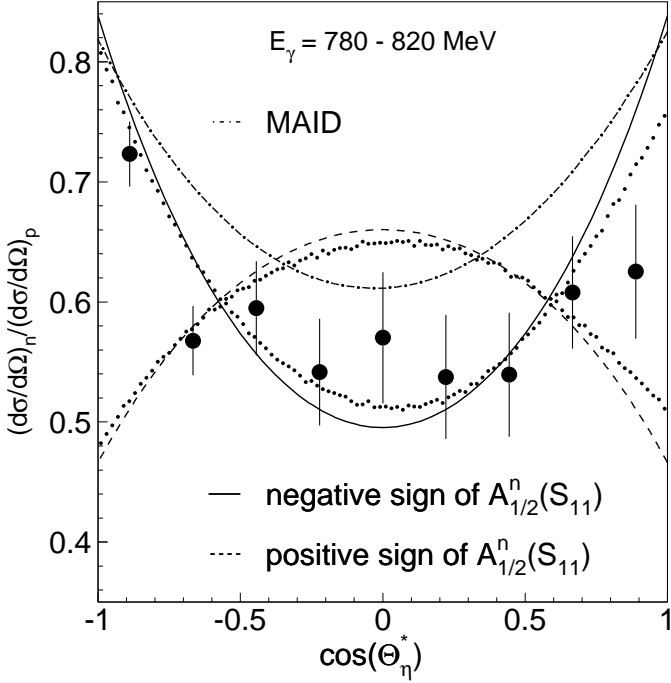}
\caption{Neutron - Proton cross section ratio as function of the $\eta$
cm angle in the excitation energy range of the S$_{11}(1535)$. The dash-dotted 
curve is the prediction of the MAID-model \cite{Eta_MAID}. The full and dashed 
lines correspond to the ratio calculated from eq. \ref{eq:dsapp} with negative 
and positive sign of $A_{1/2}^n$. The two dotted curves show the effect of
nuclear Fermi smearing and the detector acceptance on the predicted ratios.}
\label{fig:angdis}       % Give a unique label
\end{figure}
The helicity couplings of the D$_{13}$ resonance are well known from pion 
photoproduction experiments \cite{PDG}. We adopt here the results from the 
latest analysis of the GW-group \cite{Arndt} of $A_{1/2}^p(D_{13})$=(-24$\pm$2) 
(Particle Data Group \cite{PDG}: -24$\pm$9) and $A_{1/2}^n(D_{13})$=(-67$\pm$3)
(PDG: -59$\pm$9). The helicity coupling $A_{1/2}^p(S_{11})$ is taken from 
\cite{Krusche_4,Eta_MAID} as (120$\pm$20). With these parameters the constant 
${\cal{G}}(E_{\gamma})$ can be approximated at excitation energies 
around 800 MeV with eq. \ref{eq:dsapp} from the measured angular distributions 
of $p(\gamma ,\eta)p$ \cite{Krusche_1} as ${\cal{G}}(800MeV)\approx 0.2$.
Since the angular distributions for the proton have a maximum
at 90$^o$ \cite{Krusche_1}, $A_{1/2}^p(S_{11})$ is positive, and the helicity
couplings for the D$_{13}$ are both negative, we qualitatively expect a minimum
at 90$^o$ in the angular distribution of $n(\gamma ,\eta)n$ if
$A_{1/2}^n(S_{11})$ is negative and a maximum if it is positive. 

The angular dependence of the cross section ratio in the excitation region
of the S$_{11}$-resonance is shown in fig. \ref{fig:angdis} in the cm system 
of the incident photon and a nucleon at rest. In comparison to the free 
nucleon case angular distributions of quasifree $\eta$-photoproduction from 
the deuteron are only somewhat smeared out in this frame by Fermi motion 
\cite{Krusche_2} as long as the incident photon energies are not to close to 
the production threshold. Note that due to the influence of Fermi motion, the 
experiment had non-zero acceptance for all $\eta$-emission angles in this 
frame, although the lab angle of the recoil nucleons was restricted. 
The data are compared to the ratios predicted from eq. \ref{eq:dsapp} with 
$A_{1/2}^n(S_{11})=\pm 93$, corresponding to the observed cross section ratio 
in this energy region (see fig. \ref{fig:proneu}). As also demonstrated in the
figure, the effects of the momentum distribution of the nucleons and the 
limited acceptance of the detector system have only a small effect on the 
ratio of the angular distributions. It is obvious that the shape
strongly favors the negative sign (reduced $\chi^2$: 1.2 and 8.3 for negative
respectively positive sign). The data are also compared to a full 
calculation with the MAID-model \cite{Eta_MAID}, which also reproduces the 
shape with the negative sign for $A_{1/2}^n(S_{11})$. The small disagreement
on the absolute scale between data and MAID corresponds to a slightly different
absolute value of the neutron coupling. 

\subsection{Cross section ratio versus effective photon energy}
\label{sec:effratio}

Due to the momentum distribution of the bound nucleons the exclusive cross 
section ratios discussed above represent for each incident photon energy
an average over a certain range of $\sqrt{s}$ for the photon - nucleon system.
This means that when comparing to the elementary cross section ratio 
predicted by models, one must account for the Fermi smearing which could wash
out a possible energy dependence of the ratio. This effect can be
eliminated when instead of the incident photon energy the equivalent photon
energy $E_{\gamma}^{\star}$ (see eq. \ref{eq:e_eff}), reconstructed from the 
over-determined reaction kinematics, is used.   
\begin{figure}[t]
  \includegraphics[width=\columnwidth]{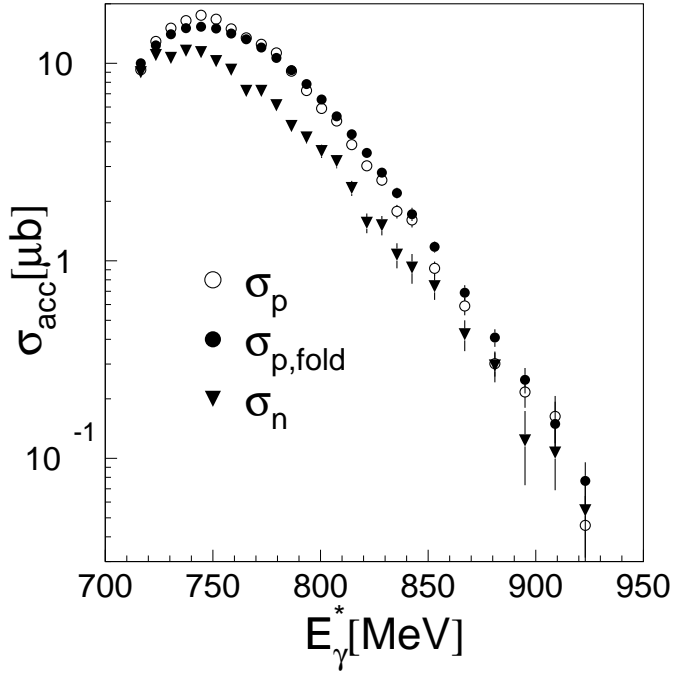}
\caption{Exclusive cross sections in the TAPS acceptance of the recoil nucleon
as function of the equivalent photon energy. ($\sigma_n$: coincident neutron,
$\sigma_p$: coincident proton, $\sigma_{p,fold}$: proton cross section folded 
with additional energy smearing).}
\label{fig:gameff}       % Give a unique label
\end{figure}

The exclusive cross sections within the TAPS acceptance as function of 
$E_{\gamma}^{\star}$ are shown in fig.
\ref{fig:gameff}. The energies of the recoil nucleons, which enter into
the calculation of the equivalent photon energy, are measured in different ways
(see sec. \ref{sec:analysis}). Consequently, systematic effects due to the 
resolution and the relative absolute calibration of the recoil nucleon 
energies must be taken into account for the extraction of the cross section 
ratio. As discussed in sec. \ref{sec:analysis} the resolution of the proton
recoil energy, determined from the energy deposited in the scintillators, is
better than for the neutron recoil energy measured via time-of-flight.
In order to avoid a systematic effect on the cross section ratio, the proton
energies were folded with an additional resolution term so that the resolution
for the recoil energies of both nucleons became equal. This was controlled via
the influence of the resolution on the shape of the missing mass spectra.
The folded proton cross section is also shown in the figure, the effect is
largest for small photon energies. 

\begin{figure}[t]
  \includegraphics[width=\columnwidth]{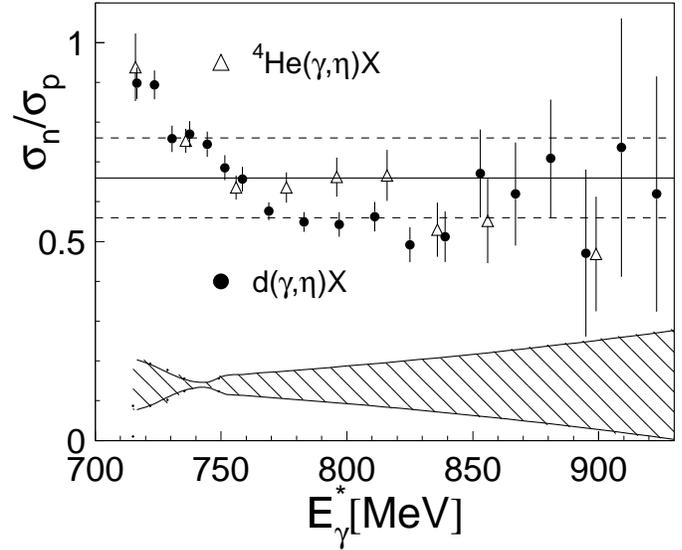}
\caption{Ratio of neutron and proton cross section as function of the equivalent
photon energy for the deuteron and for $^4$He \cite{Hejny}.
The lines indicate the average extracted from the exclusive cross sections as
function of the incident photon energy. The shaded band represents the
systematic uncertainty of the ratio due to the relative calibration of the
proton and neutron recoil energies.}
\label{fig:gameff_rat}       % Give a unique label
\end{figure}

The resulting ratio is shown in fig. \ref{fig:gameff_rat} and compared to the
measurement with the $^4$He target \cite{Hejny} which was now analyzed in the
same way. The two data sets show a very similar behavior. Due to the nuclear
re-scattering effects the ratios approach unity close to threshold. At higher
energies the ratios become almost constant although the deuteron data tend to 
somewhat lower values around photon energies of 800 MeV, where they are close 
to the lower limit extracted from fig. \ref{fig:proneu}. This behavior is 
not inconsistent with the results shown in fig. \ref{fig:proneu}.
This is demonstrated with a participant - spectator model calculation using
the cross section ratio extracted above from the equivalent photon energy.
The result for the cross section ratio as function of the incident photon
energy (dotted line in fig. \ref{fig:proneu}) agrees very well with the measured
values. On the other hand, the cross section ratio as function of the 
equivalent photon energy has an additional systematic uncertainty from the 
relative calibration of the nucleon recoil energies which can give rise to a
shift of $E_{\gamma}^{\star}$ between protons and neutrons. The possible 
size of this effect is indicated as shaded band in fig. \ref{fig:gameff_rat}. 
It was determined in the following way: the energies were artificially changed 
until a significant shift of the peak positions in the missing mass spectra was
observed. Subsequently the cross section ratio was re-analyzed with the
modified energy calibration. The uncertainty is almost negligible around 
740 MeV, where the two cross section peak, but becomes large in the regions with
a steep slope.

\section{Conclusions}
\label{sec:conclusion}

\begin{figure}[t]
  \includegraphics[width=\columnwidth]{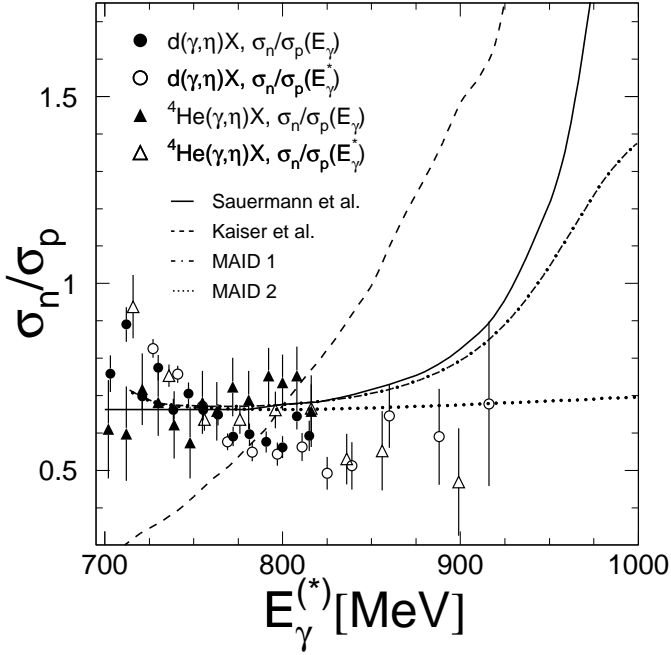}
\caption{Ratio of exclusive proton and neutron cross sections in the TAPS
acceptance. The deuteron data are compared to the $^4He$-data \cite{Hejny}
and to model predictions  from Sauermann et al. \cite{Sauermann} and
Kaiser et al. \cite{Kaiser_2}. The curves labeled MAID \cite{Eta_MAID} are the 
predictions from the full MAID model (MAID1) and a truncated version
restricted to the S$_{11}$(1535)-resonance, Born terms
and vector meson exchange (MAID2).}
\label{fig:ratall}       % Give a unique label
\end{figure}

Inclusive and exclusive $\eta$-photoproduction from deuterium has been measured.
An impulse approximation model fitted to the inclusive data yields a ratio
$\sigma_n/\sigma_p=0.67\pm0.07$ in agreement with an earlier measurement
\cite{Krusche_2}. 

Fully exclusive data from the deuteron, i.e. the coincident 
detection of the $\eta$-meson {\it and} the recoil nucleon has been obtained 
for the first time. The kinematics is over-determined for this reaction,
which allows the reconstruction of the momentum of the spectator
nucleon and of $\sqrt{s}$ of the photon - participant nucleon
system. The exclusive data was analyzed in two ways: as function of the incident
photon energy and as function of the equivalent photon energy corresponding to
the reconstructed $\sqrt{s}$. The data are compared to the results from a
similar experiment using a $^4$He target and to model predictions in fig.
\ref{fig:ratall}. 

The observed rise of the cross section ratio to unity at the breakup threshold 
for incident photon energies, respectively at the free nucleon threshold for
equivalent photon energies is understood. It is a pure nuclear effect due to 
re-scattering contributions etc., which are not related to the behavior of 
the elementary cross sections on the free nucleon. Therefore the extreme 
threshold region should not be included into the determination of the 
neutron - proton ratio for the excitation of the S$_{11}$(1535) resonance.
The ratios are almost constant for higher incident photon energies and the small
residual energy dependence is opposite for the deuteron data (slight decrease to
higher energies) and the $^4$He data \cite{Hejny} (slight increase).
As a function of the equivalent photon energy the ratios from both measurements
show a slight decrease towards higher energies in the excitation energy range 
of the S$_{11}$, which lies, however, within the systematic uncertainty of the 
data. Consequently, no final conclusion can be drawn as to whether a small 
energy dependence of the elementary cross section ratio for 
$\eta$-photoproduction on the free neutron and proton exists in the 
excitation energy range of the S$_{11}$. A comparison of the energy dependence
in the S$_{11}$ range to model predictions (see fig. \ref{fig:ratall})
clearly disfavors the interpretation of the resonance as a $K\Sigma$ bound 
state \cite{Kaiser_1,Kaiser_2}. This model reproduced the cross section for 
$p(\gamma ,\eta)p$ quite well, but predicts a strong rise of 
$\sigma_n/\sigma_p$, which is not observed.

The cross section ratios derived from the inclusive and exclusive data 
for {\em incident} photon energies in the S$_{11}$-range are compared below to 
previous results:  
 
\begin{eqnarray}
\sigma_n /\sigma_p & = & (0.67\pm 0.07) \mbox{, $^2$H, $\eta$ detected; this work} \\
 & = & (0.66\pm 0.10) \mbox{, $^2$H, $\eta$ {\it and} recoil nucleon; this work} \nonumber \\
 & = & (0.66\pm 0.07) \mbox{, $^2$H, $\eta$ detected; \cite{Krusche_2}} \nonumber \\ 
 & = & (0.68\pm 0.06) \mbox{, $^2$H, recoil nucleon detected;  \cite{Hoffmann}} \nonumber \\ 
 & = & (0.67\pm 0.07) \mbox{, $^4$He, $\eta$ detected; \cite{Hejny}} \nonumber \\ 
 & = & (0.68\pm 0.09) \mbox{, $^4$He, $\eta$ {\it and} recoil nucleon; \cite{Hejny}} \nonumber   
\end{eqnarray}
The values obtained from effective photon energies around the $S_{11}$ position
($E_{\gamma}^{\star}$=750 - 850 MeV, see fig. \ref{fig:gameff_rat}) are
close to the lower limits of the above results, but have much larger errors due
to the systematic uncertainty of the relative energy calibration:
\begin{eqnarray}
\label{res:eff}
\sigma_n /\sigma_p(E_{\gamma}^{\star}) & = & 
(0.58\pm 0.20) \mbox{, $^2$H, this work} \\
 & = & (0.62\pm 0.20) \mbox{, $^4$He, this work}\nonumber\;\;.   
\end{eqnarray}

The results from all measurements agree and their uncertainties are dominated 
by systematic effects like the absolute cross section normalizations and
the model dependence of the extraction from inclusive data. This systematic 
uncertainties are largely independent since the experiments include different 
target nuclei, different extraction methods of the ratio, and different 
experimental setups, so that an overall average of the results is meaningful:  
\begin{eqnarray}
\langle \sigma_n /\sigma_p \rangle & = & (0.67\pm 0.03) \\
|A_{1/2}^n|/|A_{1/2}^p| & = & (0.819\pm 0.018) \\
 A_{1/2}^{IV}/A_{1/2}^{IS} & = & (10.0\pm 0.7) \\ 
 A_{1/2}^{IS}/A_{1/2}^{p} & = & (0.09\pm 0.01) \nonumber
\end{eqnarray}

Inclusion or not of the less precise results obtained from the analysis
of the effective photon energies ( eq. \ref{res:eff}) leaves the overall result 
unchanged. This precise value of the cross section ratio can then be used for 
the extraction of the ratio of the electromagnetic helicity couplings 
$A_{1/2}^n$ and $A_{1/2}^p$ for the excitation of the S$_{11}$(1535) resonance 
on the neutron and the proton. Here it is assumed that in the energy range of 
interest $\eta$-photoproduction is so strongly dominated by the excitation 
of the S$_{11}$ resonance that the sum of all other contributions can be safely
neglected \cite{Krusche_4}. The extraction of the isospin structure of the
helicity amplitude, i.e. the ratio $A_{1/2}^{IV}/A_{1/2}^{IS}$ or
$A_{1/2}^{IS}/A_{1/2}^{p}$ furthermore requires the knowledge 
of the relative phase between the proton and the neutron amplitude. The results
from coherent $\eta$-photoproduction from the deuteron \cite{Hoffmann,Weiss}
and the behavior of the ratio of the angular distributions for quasifree 
$\eta$-production from the proton and the neutron investigated here,
have shown, that the reaction on the nucleon must be dominantly isovector.
Under this assumption a 9\% isoscalar contribution to the proton amplitude is
determined. The lowest value for the proton - neutron ratio in the S$_{11}$ 
range is 0.55 for equivalent photon energies around 800 MeV in the 
present experiment. Even if we use this value, the isoscalar component would 
only rise to 13\%.   
As discussed in \cite{Weiss} such a small isoscalar amplitude is not
in agreement with models reproducing the measured coherent cross section
which require an isoscalar admixture of roughly 20\%. The origin of this
discrepancy is not yet understood. It was however recently suggested by Ritz 
and Arenh\"ovel \cite{Ritz} that the proton and neutron amplitudes could have 
an unexpected large {\it complex} relative phase. In this case the amplitude 
ratio $|A_{1/2}^n|/|A_{1/2}^p| = (0.819\pm 0.018)$ could be consistent with a
20\% isoscalar admixture.

At higher photon energies a rise of the neutro-proton ratio is predicted by
models. This is shown in fig. \ref{fig:ratall}. The full curve is a prediction
from a coupled channel model by Sauermann et al. \cite{Sauermann}. The
dash-dotted and dotted curves are predictions from the MAID model
\cite{Eta_MAID} for the full model and for a truncated model which only takes
into account the excitation of the S$_{11}$(1535) resonance, Born terms and
vector meson exchange. The predicted rise at the higher photon energies is
assigned to the excitation of higher lying nucleon resonances. The measured
cross section ratio does not yet show any rise to higher energies, but it is
evident that higher incident photon energies must be used to test this
prediction.

~\\
{\bf Acknowledgments}

We wish to acknowledge the outstanding support of the accelerator group of
MAMI, as well as many other scientists and technicians of the Institut f\"ur
Kernphysik at the University of Mainz.
Stimulating discussions with A. Arenh\"ovel and A. Fix are gratefully
acknowledged. 
This work was supported by Deutsche Forschungsgemeinschaft (SFB 201), the UK 
Engineering and Physical Science Research Council and the Swiss National 
Science Foundation.

\end{document}